\documentclass[AMA,Times1COL]{WileyNJDv5}
\makeatletter
\let\oddfoot@titlepage@info\relax
\makeatother

\articletype{Original Research}%

\received{Date Month Year}
\revised{Date Month Year}
\accepted{Date Month Year}
\journal{Journal}
\volume{00}
\copyyear{2023}
\startpage{1}

\begin{document}

\title{\textbf{Combined Diffusion-Relaxation MRI to Assess Muscle
Microstructure and Composition}
}

\author[1]{Matteo Figini}
\author[2,3,5]{Paddy J. Slator}
\author[4]{Valeria E. Contarino}
\author[1]{Eleftheria Panagiotaki}
\author[5]{Giovanna Rizzo}
\author[5]{Alfonso Mastropietro}

\authormark{FIGINI \textsc{et al.}}
\titlemark{Combined Diffusion‐Relaxation MRI to Assess Muscle
Microstructure and Composition}

\address[1]{\orgdiv{Hawkes Institute and Department of Computer Science}, \orgname{University College London}, \orgaddress{\state{London}, \country{United Kingdom}}}

\address[2]{\orgdiv{Cardiff University Brain Research Imaging Centre}, \orgname{School of Psychology}, \orgaddress{\state{Cardiff}, \country{United Kingdom}}}

\address[3]{\orgdiv{School of Computer Science and Informatics}, \orgname{Cardiff University}, \orgaddress{\state{Cardiff}, \country{United Kingdom}}}

\address[4]{\orgdiv{Neuroradiology Unit}, \orgname{Fondazione IRCCS Ca'Granda Ospedale Maggiore Policlinico}, \orgaddress{\state{Milano}, \country{Italy}}}

\address[5]{\orgdiv{Istituto di Sistemi e Tecnologie Industriali Intelligenti per il Manifatturiero Avanzato}, \orgname{Consiglio Nazionale delle Ricerche}, \orgaddress{\state{Milano}, \country{Italy}}}

\corres{Corresponding author Matteo Figini. \email{m.figini@ucl.ac.uk}}


\abstract[Abstract]{Quantifying muscle tissue properties is crucial for understanding pathophysiological changes occurring in skeletal muscle (SM). In particular, $\mathrm{T}_2$ relaxation and diffusion MRI (dMRI) are  promising techniques. However, typical methods measure $\mathrm{T}_2$ and diffusion separately, making them less specific to microstructure than emerging combined diffusion-relaxation techniques. Here we demonstrate a combined diffusion-relaxation MRI approach for disentangling $\mathrm{T}_2$ and diffusivity properties in SM.
A unified diffusion–relaxation acquisition was implemented on a 3 T scanner, combining six b-values (15–600 s/mm²) and four echo times (TEs = 50–90 ms) within a 12-min single-slice protocol. Five healthy participants (32 ± 6 years) were enrolled. Data were analysed with six microstructural diffusion and diffusion-relaxation models. Mean parameter values were extracted from manually segmented calf muscles—Tibialis Anterior (TIB), Peroneus (PER), Gastrocnemius Medialis (GM), Gastrocnemius Lateralis (GL), and Soleus (SOL).
Models neglecting $\mathrm{T}_2$ relaxation showed strong TE dependence: mean diffusivity (MD) decreased by up to 47\%, fractional anisotropy (FA) increased by up to 75\%, and vascular fraction ($\mathrm{f}_V$) was overestimated by up to 297\% when TE increased from 50 to 90 ms. Diffusion–relaxation models produced TE-independent estimates. Across these models, MD and $D_t$ were higher in PER, GM, GL, and SOL than in TIB (differences up to +14\%), whereas FA was lowest in SOL (up to -42\%). Tissue and vascular relaxation times ranged 31–36 ms ($T_{2t}$) and 66–86 ms ($T_{2v}$), respectively. Simulations confirmed improved accuracy for $\mathrm{f}_V$ estimation (r = 0.95; RMSE = 0.03) and reduced TE-related bias.
Combined diffusion–relaxation MRI provides robust, TE-independent estimates of muscle microstructural and perfusion-related biomarkers. The quantitative improvements observed — particularly in the estimation of $\mathrm{f}_V$ - show its potential to provide non-invasive biomarkers for the assessment of muscle physiology, exercise adaptation, rehabilitation, and neuromuscular pathology.
}


\jnlcitation{\cname{%
\author{Figini M},
\author{Slator P},
\author{Contarino V},
\author{Panagiotaki L},
\author{Rizzo G}, and
\author{Mastropietro A}}.
\ctitle{Combined Diffusion‐Relaxation MRI to Assess Muscle
Microstructure and Composition.} \cjournal{\it NMR Biomed} \cvol{2021;00(00):1--18}.}

\maketitle

\renewcommand\thefootnote{}
\footnotetext{\textbf{Abbreviations:} MRI, magnetic resonance imaging; dMRI, diffusion MRI; IVIM, intravoxel incoherent motion; DTI, diffusion tensor imaging}

\renewcommand\thefootnote{\fnsymbol{footnote}}
\setcounter{footnote}{1}

\section{Introduction}\label{sec1}

MRI enables the non-invasive assessment of skeletal muscle (SM) properties, reducing the need for invasive biopsies. Non-invasive techniques are in high clinical demand for quantifying the effects of physical activity, rehabilitation treatments, aging, and pharmacological interventions on muscle tissue \cite{berry2024voxels}. MRI is proving to be particularly valuable in the diagnosis and follow-up of muscular dystrophies \cite{diaz2015muscle, mastropietro2024classification}, as well as in evaluating sports injuries \cite{isern2024muscle} and assessing the effects of treatments and rehabilitation \cite{mastropietro2025multiparametric}. A key advantage of MRI is its ability to provide spatial maps of quantitative metrics directly linked to muscle structure and function. 

Diffusion MRI (dMRI) is sensitive to changes in the microstructural properties of SM and has been used to investigate SM perfusion \cite{adelnia2023mri}, microstructure \cite{oudeman2016techniques} and fiber composition \cite{scheel2013fiber}. Due to the highly anisotropic nature of SM, models such as diffusion tensor imaging (DTI) are well suited and have been widely applied, making it a valuable tool for i) fiber orientation mapping to estimate the orientation and alignment of muscle fibers; ii) detecting  microstructural changes  due to disease, injury, or exercise; iii) functional assessment to investigate alterations, for example during and after physical activity. While DTI effectively characterizes anisotropic diffusion, its assumptions oversimplify complex microstructural environments.

The intravoxel incoherent motion (IVIM) model, which separates perfusion- and tissue-related signal components, has also been widely used in dMRI of SM \cite{englund2022intravoxel}. Its sensitivity to changes in both diffusion and perfusion makes it useful for investigating physiological processes such as exercise-induced changes to track perfusion and tissue diffusivity alterations during and after exercise \cite{mastropietro2018triggered}; for studying pathological conditions to identify abnormalities in muscle perfusion and diffusion caused by neuromuscular diseases \cite{ran2021diagnostic}; for rehabilitation assessment to monitor changes in muscle function and perfusion after therapy or injury \cite{shahidi2021intravoxel, shu2021acute}.

Some approaches have combined elements of IVIM and DTI, enabling models that separate perfusion and diffusion whilst also accounting for anisotropy. These \textit{anisotropic} IVIM models have been applied to muscle \cite{karampinos2010intravoxel}, brain \cite{finkenstaedt2017ivim}, heart \cite{zhang2021investigation}, kidney \cite{liu2018renal}, placenta \cite{slator2018placenta}, and peripheral nerves \cite{merchant2022simultaneous}. These models can enable new imaging biomarkers, such as separate perfusion- and diffusion-related fibre direction. These extensions typically work by replacing the isotropic assumption of the two compartments with anisotropic models such as the diffusion tensor or its simplified version, the Zeppelin, i.e. an axially symmetric tensor. A further extension entails the addition of a third compartment with restricted diffusion, as in the Vascular, Extracellular, and Restricted Diffusion for Cytometry in Tumors (VERDICT) framework, which has been extensively applied to body and brain tumor imaging \cite{panagiotaki2015verdict, singh2023prostate, figini2023comprehensive}.

Relaxation MRI is also sensitive to SM tissue microstructure, with $T_1$ and $\mathrm{T}_2$ providing some sensitivity to the composition of fast-twitch and slow-twitch fibres \cite{bonny1998characterization}. Additionally, $\mathrm{T}_2$ has shown sensitivity to changes due to exercise \cite{patten2003t2} and disease status as in muscular dystrophy \cite{wokke2016t2, mankodi2017skeletal}.

However, measuring diffusion and relaxation properties in separate experiments oversimplifies tissue microstructure, and complex microstructural environments can only be comprehensively characterised by combined diffusion-relaxation MRI \cite{slator2021combined}. 

This is particularly relevant for  SM, where hierarchical organisation impacts diffusion across various length scales, and there is added complexity due to diverse relaxation and diffusion properties in different muscle structures and conditions \cite{bonny1998characterization, ochi2016differences, cameron2023age}. 

Moving to combined diffusion-relaxation data necessitates corresponding extensions to the models, for example although the IVIM model and its extensions described above provide a robust framework, they assume that both perfusion and diffusion compartments share the same $\mathrm{T}_2$ relaxation time, which may be an oversimplified hypothesis in complex tissue environments. Extensions such as the $\mathrm{T}_2$-IVIM model \cite{jerome2016extended} address this limitation by allowing for distinct $\mathrm{T}_2$ values in each compartment, enhancing parameter estimation accuracy.

While integrated diffusion–relaxometry models have been successfully applied in the brain, placenta, cancer, and other organs, their use in SM remains unexplored, with previous studies limited to separate relaxation and diffusion MRI acquisitions \cite{li2014multi, stouge2020mri, otto2020quantitative}. Here, we present the first in vivo implementation of truly integrated $\mathrm{T}_2$‑diffusion MRI in SM, leveraging a unified acquisition and modeling framework. Our approach simultaneously estimates diffusion and relaxation parameters, enabling reliable quantification of microstructural changes in muscle tissue.

\begin{figure*}
\centerline{\includegraphics[width=\textwidth]{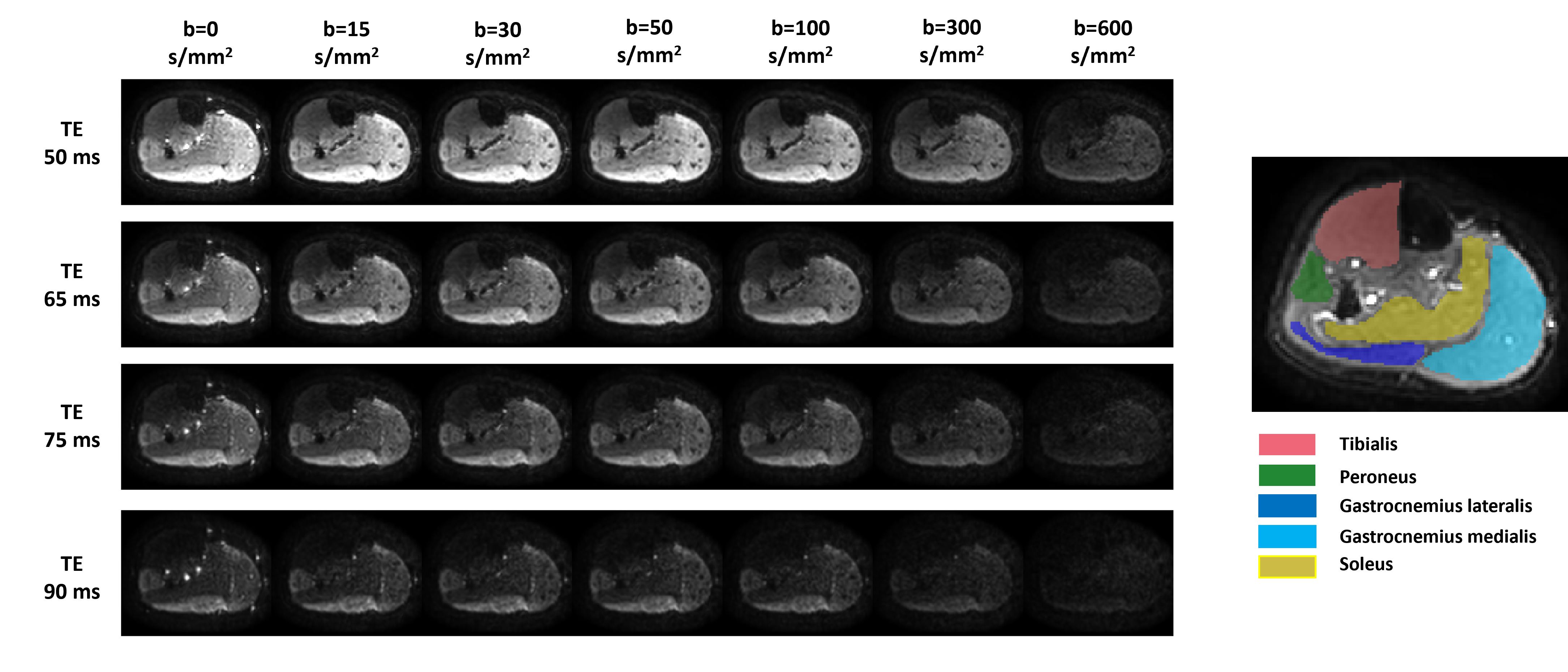}}
\caption{Representative diffusion-weighted images of the calf acquired at varying b-values and echo times (TEs) for a single diffusion direction. The panel on the right shows manually segmented ROIs for five calf muscles: Tibialis (TIB), Peroneus (PER), Gastrocnemius Lateralis (GL), Gastrocnemius Medialis (GM), and Soleus (SOL).\label{fig1}}
\end{figure*}

\section{Methods}\label{sec2}

\subsection{Theoretical Models Description}

The modelling approach explored in this study builds upon the classical IVIM model \cite{le1986mr}, which distinguishes diffusion and perfusion (or pseudo-diffusion) in isotropic compartments. We then extended the IVIM model by incorporating anisotropic diffusion and $\mathrm{T}_2$ relaxation effects, enabling a comprehensive analysis of the SM microstructure. The following section outlines the theoretical framework underpinning our modeling approach. All included models and estimated parameters are listed in table \ref{table1}. 

\subsubsection{Ball-Ball and Ball-Ball-T2}
The IVIM model describes the MRI signal as the weighted sum of two isotropic compartments, i.e. each compartment is a “Ball” in the notation of \cite{panagiotaki2012nimg}; one representing tissue diffusion and another representing pseudo-diffusion from blood flow in capillaries.

\begin{align}\label{eq1}
S(b) =  S_0 \left[ f_v \cdot e^{-b \cdot D_v} + (1-f_v) e^{-b \cdot D_t} \right]
\end{align}

where $f_v$ is the signal fraction of the vascular (pseudo-diffusion) compartment, $D_v$ is the pseudo-diffusivity in the vascular compartment and $D_t$ is the diffusivity in the tissue. From here onwards we refer to this model as "Ball-Ball".

However, the IVIM model assumes that the perfusion- and tissue-related compartments both have the same $\mathrm{T}_2$ value. In \cite{jerome2016extended} the authors extended this account for different $\mathrm{T}_2$ values:

\begin{align}\label{eq2}
S(b, T_E) =  S_0 \left[ f_v \cdot e^{-b \cdot D_v} e^{-\frac{T_E}{T_{2v}}} + (1-f_v) \cdot e^{-b \cdot D_t} e^{-\frac{T_E}{T_{2t}}} \right]
\end{align}

where ${T_{2v}}$ and ${T_{2t}}$ are the ${T_2}$ relaxation times in the vessels and tissue respectively. We refer to this model as "Ball-Ball-$\mathrm{T}_2$".

Furthermore, the authors of \cite{jerome2016extended} demonstrated that if the perfusion fraction is estimated using the standard IVIM model, then, if  the perfusion- and tissue-related compartments have different $\mathrm{T}_2$ values, this affects the estimated perfusion fraction as follows:

\begin{align}\label{eq3}
f_{\text{est}}(T_E) = \frac{f_v \cdot e^-{\frac{T_E}{T_{2v}}}}{f_v \cdot e^-{\frac{T_E}{T_{2v}}} + (1-f_v) \cdot e^-{\frac{T_E}{T_{2t}}}}
\end{align}

where $f_{est}$ represents the estimated perfusion fraction and $f_v$ denotes the true underlying perfusion fraction.

\subsubsection{Ball-Zeppelin and Ball-Zeppelin-T2}

Several extensions of the IVIM model have been developed to quantify anisotropy, and these have been applied to muscle. Here we use the "Ball-Zeppelin" model, an extension of the standard IVIM model, where the diffusion compartment is replaced by a Zeppelin (using the terminology for diffusion compartments from \cite{panagiotaki2012nimg}), i.e. an axially-symmetric diffusion tensor. The diffusion signal decay described by a diffusion tensor is:

\begin{align}\label{eq4}
S(b) = S_0 \cdot e^{-b \mathbf{g}_k^T \textbf{D} \mathbf{g}_k}
\end{align}

where \textbf{$g_k$} is the unit vector for the kth diffusion-sensitizing gradient and \textbf{D} is a diffusion tensor(\cite{basser1994dti}), with the principal eigenvector (main diffusion direction) defined by angles $\theta$ and $\phi$. 
The assumption of axial symmetry means that diffusivity in any direction perpendicular to the main eigenvector has the same value $D_{\perp}$ (Radial Diffusivity), while diffusivity along it is defined as $D_{\parallel}$ (Axial Diffusivity). So in the frame of reference of its eigenvectors, \textbf{D} is

\begin{align}\label{eq5}
\textbf{D} = \begin{bmatrix}
D_{\parallel} & 0 & 0\\
0 & D_{\perp} & 0\\
0 & 0 & D_{\perp}
\end{bmatrix}
\end{align}

So the Ball-Zeppelin model becomes:

\begin{align}\label{eq6}
S(b) = S_0 \left[ f_v \cdot e^{-b D_v} + (1 - f_v) \cdot e^{-b \cdot \mathbf{g}_k^T \textbf{D} \mathbf{g}_k} \right]
\end{align}

The Ball-Zeppelin model can be further extended by including $\mathrm{T}_2$ relaxation in the same way as above: 

\begin{align}\label{eq7}
S(b,T_E) = S_0 \left[ f_v \cdot e^{-\frac{T_E}{T_{2v}}} e^{-b \cdot D_v} 
+ (1 - f_v) \cdot e^{-\frac{T_E}{T_{2t}}} e^{-b \mathbf{g}_k^T \textbf{D} \mathbf{g}_k} \right]
\end{align}

We refer to this as the "Ball-Zeppelin-$\mathrm{T}_2$" model in the following.

\subsubsection{Zeppelin and Zeppelin-T2}
We also considered a simple Zeppelin, as a reference for the most common clinical standard (DTI), and a Zeppelin-$\mathrm{T}_2$ model:

\begin{align}\label{eq8}
S(b, T_E) = S_0 \cdot e^{-\frac{T_E}{T_{2t}}} \cdot e^{-b \mathbf{g}_k^T \textbf{D} \mathbf{g}_k}
\end{align}

Note that these models do not consider the pseudo-diffusion component of the signal, so they are not appropriate at low b-values.

\begin{table}
    \begin{center}
     \caption{List of included model and of the free parameter estimated by each of them; see the text for the mathematical formulations.}
    \label{table1}
    \begingroup
    \def\arraystretch{1.5}
     \begin{tabular}{l|l}
        \textbf{Model} & \textbf{Free parameters}\\
        \toprule
         Ball-Ball & $f_v$, $D_v$, $D_t$ \\
         Ball-Ball-$\mathrm{T}_2$ & $f_v$, $D_v$, $D_t$, $T_{2v}$, $T_{2t}$ \\
         Ball-Zeppelin & $f_v$, $D_v$, $D_{\parallel}$, $D_{\perp}$, $\theta$, $\phi$ \\
         Ball-Zeppelin-$\mathrm{T}_2$ & $f_v$, $D_v$, $D_{\parallel}$, $D_{\perp}$, $\theta$, $\phi$, $T_{2v}$, $T_{2t}$ \\
         Zeppelin & $D_{\parallel}$, $D_{\perp}$, $\theta$, $\phi$\\
         Zeppelin-$\mathrm{T}_2$ & $D_{\parallel}$, $D_{\perp}$, $\theta$, $\phi$, $\mathrm{T}_2$\\
         \bottomrule
    \end{tabular}
    \endgroup
    \end{center}
        See table \ref{table2} for the meaning of each parameter
\end{table}

\begin{table}
    \begin{center}
     \caption{Physically meaningful range for each model parameter, used for sampling the ground truth values in the training set}
    \label{table2}
    \begingroup
    \def\arraystretch{1.5}

     \begin{tabular}{p{0.7cm}|p{3.2 cm}|p{1.5cm}|p{1.6cm}}
        \textbf{Param} & \textbf{meaning} & \textbf{min} & \textbf{max}\\
        \toprule
        $f_v$ & vascular fraction & 0 & 1 *\\
        $D_v$ & pseudo-diffusivity in the vascular compartment & 10 $\mu m^2/ms$ & 100 $\mu m^2/ms$\\
        $D_t$ & tissue diffusivity & 0.1 $\mu m^2/ms$ & 3 $\mu m^2/ms$ \\
        $D_{\parallel}$ & tissue axial diffusivity & 0.1 $\mu m^2/ms$ & 3 $\mu m^2/ms$\\
        $D_{\perp}$ & tissue radial diffusivity & 0.1 $D_{\parallel}$ & $D_{\parallel}$ **\\
        $\theta$ & polar angle identifying the Zeppelin principal direction & 0 & $\pi$\\
        $\phi$ & azimuthal angle identifying the Zeppelin principal direction & 0 & $\pi/2$\\
        $T_{2v}$ & vascular $\mathrm{T}_2$ relaxation time & 10 ms & 200 ms\\
        $T_{2t}$ & tissue $\mathrm{T}_2$ relaxation time & 10 ms & 200 ms\\
        $\mathrm{T}_2$ & global $\mathrm{T}_2$ relaxation time & 10 ms & 200 ms \\
        \bottomrule
    \end{tabular}
    \endgroup
    \end{center}
    Notes: *In our current implementation, we estimate the tissue fraction $f_t$ rather than the vascular fraction, and derive $f_v = 1-f_t$. Also, we use a squared cosine transformation to ensure that fractions are between 0 and 1 at test stage. Thus, we use an intermediate parameter p1, sampled between 0 and $\pi$ in the training set, and derive $f_t=cos(p1)^2$.\\
    **To ensure that $D_{\perp} \leq D_{\parallel}$, we use an intermediate parameter p2, sampled between 0 and 1 in the training set, and derive $D_{\perp}=p2 \cdot D_{\parallel}$
    
\end{table}
    
\begin{figure*}
\centerline{\includegraphics[width=0.7\textwidth]{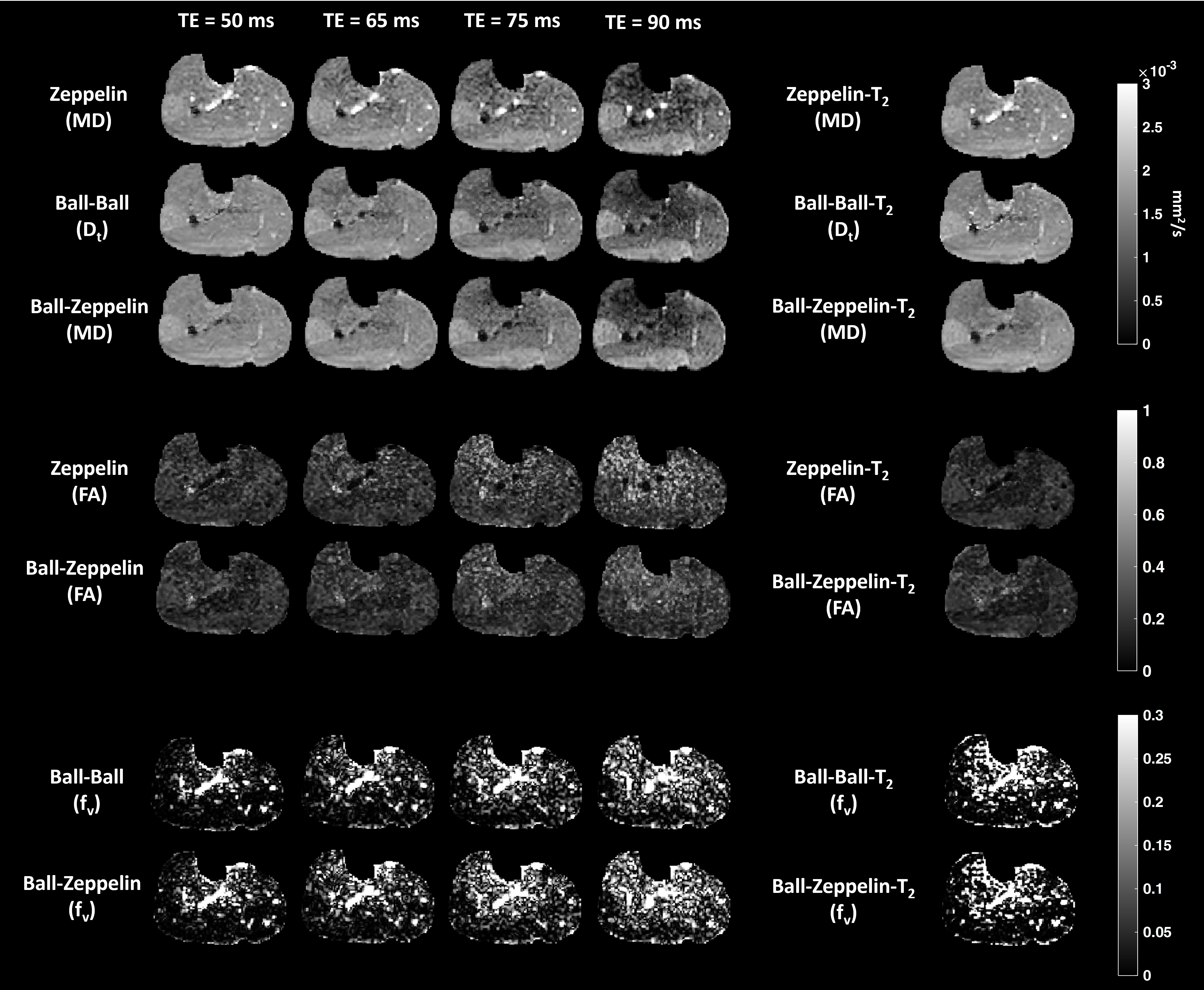}}
\caption{Representative parametric maps of MD, FA, and $f_v$ derived from models that either incorporate or neglect $\mathrm{T}_2$ relaxation. The maps illustrate how parameter estimates are affected by TE when $\mathrm{T}_2$ effects are not modelled, resulting in visible TE-dependent variations. In contrast, models that account for $\mathrm{T}_2$ provide parameters estimations not affected by $\mathrm{T}_2$ effects .\label{fig2}}
\end{figure*}

\subsection{MRI Protocol}
Five healthy participants (2 males, 3 females; age: 32 $\pm$ 6 years) were enrolled in this study after providing written informed consent. The study protocol was reviewed and approved by the CNR Ethics Committee (Protocol number: 0142383; 29/04/2024), and all procedures were conducted in accordance with the Declaration of Helsinki and relevant national guidelines. Before MRI scans, volunteers performed repetitive tiptoe standing exercises for up to 10 minutes or till exhaustion. MRI data were acquired on a 3T Philips Achieva scanner using a combined $\mathrm{T}_2$-diffusion protocol with simultaneous variation of diffusion encoding and echo time (TE). Diffusion-weighted images were obtained for $b = 15, 30, 50, 100, 300, 600 \, \text{s/mm}^2$ and $TE = 50, 65, 75, 90 \, \text{ms}$. For each $b$-$TE$ pair, six diffusion gradient directions were used, resulting in 144 diffusion-weighted images. Additionally, a pair of $b = 0$ images with opposite phase encoding directions was acquired at each TE. TR was set to 1000 ms A single slice (thickness: 5 mm; matrix size: 352 $\times$ 352; in-plane resolution: 1.2 mm) was obtained from the dominant calf. The total acquisition time was 12 minutes. An example of the acquired images for one diffusion gradient direction is shown in Figure\ref{fig1}.

\subsection{Image Preprocessing}
MR images were denoised and Gibbs ringing artefacts removed using \textit{dwidenoise} and \textit{mrdegibbs} in MRtrix \cite{veraart2016denoising,kellner2016gibbs}. Susceptibility, eddy current-induced distortions and subject movements were corrected using \textit{topup} and \textit{eddy} and all volumes were coregistered to the reference ($b = 0$, $TE=50$ ms) using flirt in FSL \cite{andersson2003correct,smith2004advances}.

\subsection{Quantitative Parameter Extraction}

For each microstructure model, a Multi-Layer Perceptron (MLP) was trained to learn the relationship between signals and model parameters using the approach in \cite{figini2024ismrm,palombo2023scirep,sen2022diagnostics}. Briefly, one million combinations of model parameters were randomly sampled from uniform distributions within physically-meaningful ranges (see \ref{table2}). Synthetic signals were generated from these ground-truth parameter combinations using the formulas above (equations \ref{eq1}, \ref{eq2}, \ref{eq4}, \ref{eq6}, \ref{eq7}, \ref{eq8}) with a correction factor accounting for normalisation. Rician noise was added to obtain an SNR value uniformly sampled between 10 and 50 on the first b=0 volume; this range of SNRs was designed to mimic the average values and spatial variability in-vivo. For models with $\mathrm{T}_2$, signals were synthesised according to the full acquisition protocol resulting in 148 data points; for models without $\mathrm{T}_2$, signals were synthesized for a single TE in the acquisition (37 data points). For the Zeppelin and $\mathrm{T}_2$-Zeppelin models, only $b = 600 \text{ s/mm}^2$ was considered (7 and 28 data points respectively).
The MLP was implemented in scikit-learn with 3 hidden layers of size 150, ReLu activation function, using an Adam optimizer with initial learning rate 0.001; the loss function was the mean square error between predicted and ground-truth parameters.
At test stage, the trained MLPs were applied to the normalised signal in each voxel of each subject to obtain microstructure parameter estimations. For models without $\mathrm{T}_2$ this was repeated separately for each TE, and for the Zeppelin and Zeppelin-$\mathrm{T}_2$ models only signals at $b = 600 \text{ s/mm}^2$ were considered.
For the models including Zeppelin, the Mean Diffusivity (MD) and Fractional Anisotropy (FA) (\cite{basser1996mdfa}) were computed as:

\begin{align}\label{eq9}
MD = \frac{D_{\parallel} + 2 \cdot D_{\perp}}{3}\end{align}

\begin{align}\label{eq10}
FA = \sqrt{\frac{3}{2}} \cdot \sqrt{\frac{(D_{\parallel}-MD)^2+2(D_{\perp}-MD)^2}{D_{\parallel}^2+2 \cdot D_{\perp}^2}}
\end{align}
 
\subsection{ROI analysis}

The average value of each quantitative parameter was computed within five Regions of Interest (ROIs), manually delineated on $b0$ images at the lowest TE for each subject, in the Tibialis (TIB), Peroneus (PER), Gastrocnemius Lateralis (GL), Gastrocnemius Medialis (GM), and Soleus (SOL) muscles (as illustrated in Figure\ref{fig1}). 

\begin{figure*}
\centerline{\includegraphics [width=\textwidth] {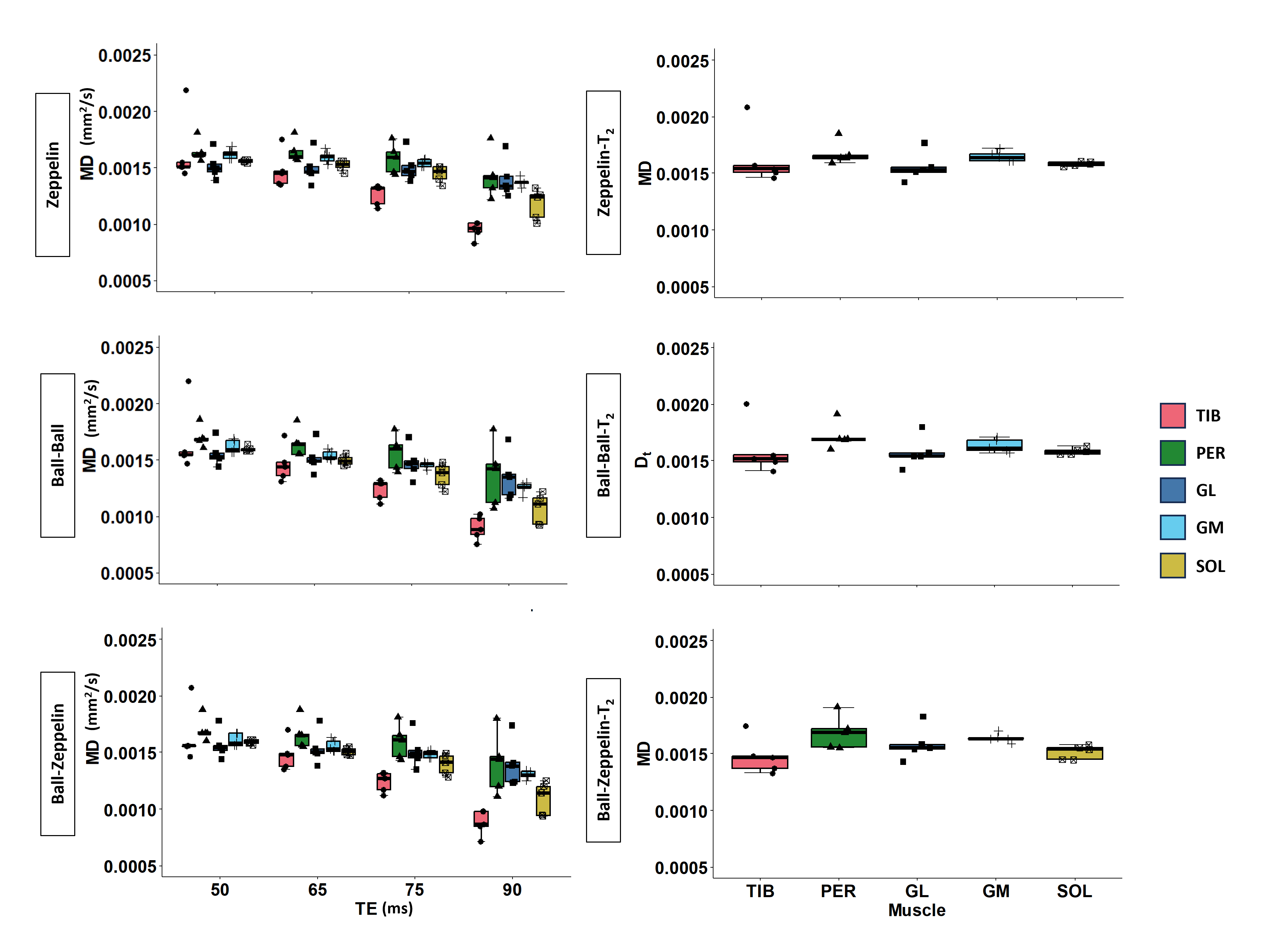}}
\caption{Boxplots of ROI-averaged mean diffusivity values across five subjects for all calf muscles. The upper panel shows results from models that do not incorporate $\mathrm{T}_2$ relaxation, illustrating the dependence of diffusivity estimates on TE. The lower panel displays TE-independent diffusivity estimates obtained from models that account for $\mathrm{T}_2$ relaxation.\label{fig3}}
\end{figure*}

\section{Results}\label{sec3}

Figure \ref{fig2} shows maps of diffusivity (MD or $D_t$), FA and $f_v$ for one representative subject, estimated using models both with and without relaxation. The diffusivity maps reveal higher values in the PER and GM muscles compared, and GL in particular with high TEs, to TIB and SOL. In contrast, FA appears slightly higher in the TIB and PER muscles relative to the others. For $f_v$,  GL and PER exhibit lower values compared to the remaining muscle groups.

\subsection{Diffusion Models}

As shown in Figure \ref{fig2}, the contrast of all the maps extracted from diffusion models not accounting for $\mathrm{T}_2$ shows a clear dependence on TE. This TE dependence is confirmed by the quantitative analysis. Figures 3-5 show bloxplots of the ROI averages of the same parameters as in figure \ref{fig2} across all subjects. MD and $D_t$ (figure \ref{fig3}) decrease with TE. As an example, in the TIB muscle, increasing TE from 50 ms to 90 ms reduces the estimated MD by 42\% with the Zeppelin model, 46\% with the Ball–Ball model, and 47\% with the Ball–Zeppelin model. 
A very similar decreasing trend with TE was also observed for $D_{\parallel}$ and $D_{\perp}$ in the models containing a Zeppelin compartment (data not shown).

By contrast, both FA (figure \ref{fig4}) and $f_v$ (figure \ref{fig5}) tend to increase with TE. Increasing TE from 50 to 90 ms can result in maximum FA increases of 75\% in the TIB muscle when using the Zeppelin model and of 75\% in the SOL muscle with the Ball–Zeppelin model. $f_v$ increases reach up to 297\% in the GL muscle with the Ball–Ball model, and up to 244\% in the same muscle when using the Ball–Zeppelin model.

The relationship between parameter values across different muscles can also vary across all models, as the absolute values for each muscle are differently affected by TE (see figures \ref{fig3}-\ref{fig5}). Differences between muscles were significant in most of the conditions explored by the models and in particular with long TEs. 

MD and $D_t$ were generally higher in PER, GL, GM, and SOL than in TIB, particularly at longer TEs; however, the absolute values and the relative differences between muscles were substantially affected by TE (figure \ref{fig3}). 

For example, using the Zeppelin model, pairwise differences between TIB and the other muscles ranged from approximately -0.1\% to +7.7\% at TE = 50 ms and from -50.5\% to -24.2\% at TE = 90 ms. With the Ball–Ball model, $D_t$ differences ranged from -2.2\% to +6.6\% at TE = 50 ms and from -53.0\% to -19.5\% at TE = 90 ms. Similarly, with the Ball–Zeppelin model, MD differences ranged from -3.5\% to +4.4\% at TE = 50 ms and from -60.1\% to -25.0\% at TE = 90 ms.

Considering FA, inter-muscle differences were also affected by TE. Using the Zeppelin model, pairwise FA differences between TIB and the other muscles ranged from -10.5\% to +19.7\% at TE = 50 ms, from +2.7\% to +26.2\% at TE = 65 ms, from +10.9\% to +23.4\% at TE = 75 ms, and from +20.4\% to +33.2\% at TE = 90 ms. With the Ball–Zeppelin model, FA differences ranged from -14.0\% to +35.8\% at TE = 50 ms, from -6.6\% to +34.2\% at TE = 65 ms, from +4.9\% to +30.8\% at TE = 75 ms, and from +21.2\% to +36.3\% at TE = 90 ms

As for $f_v$, TIB generally showed higher values than the other muscles - and in particular compared to PER and GM - although the magnitude of the differences varied across TE. With the Ball–Ball model, pairwise $\mathrm{f}_V$ differences between TIB and the other muscles ranged from +19.9\% to +76.1\% at TE = 50 ms, from 0.0\% to +65.3\% at TE = 65 ms, from +6.0\% to +59.8\% at TE = 75 ms, and from -1.7\% to +54.1\% at TE = 90 ms. With the Ball–Zeppelin model, the corresponding ranges were +25.1\% to +75.3\%, +2.2\% to +64.8\%, +8.5\% to +60.2\%, and +0.2\% to +55.5\%, respectively. 

\subsection{Diffusion-Relaxation Models}

\begin{figure*}
\centerline{\includegraphics[width=1\textwidth]{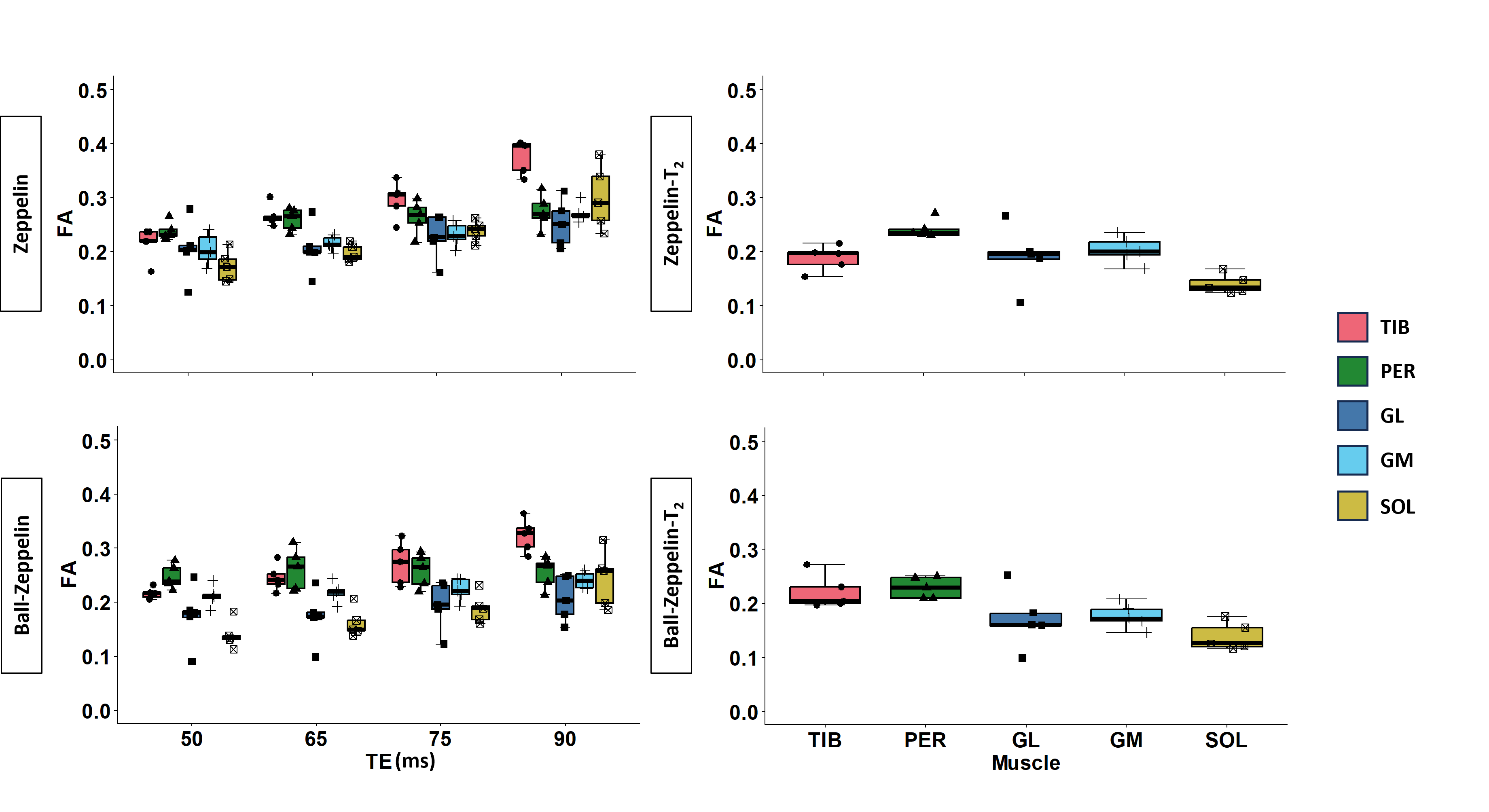}}
\caption{FA across models with and without $\mathrm{T}_2$ relaxation. Boxplots of ROI-averaged FA values from all considered diffusion models, grouped by muscle. The left panels include models without $\mathrm{T}_2$ relaxation, showing variability with TE, while the right panels include models that account for $\mathrm{T}_2$ effects.\label{fig4}}
\end{figure*}

\begin{figure*}
\centerline{\includegraphics[width=1\textwidth]{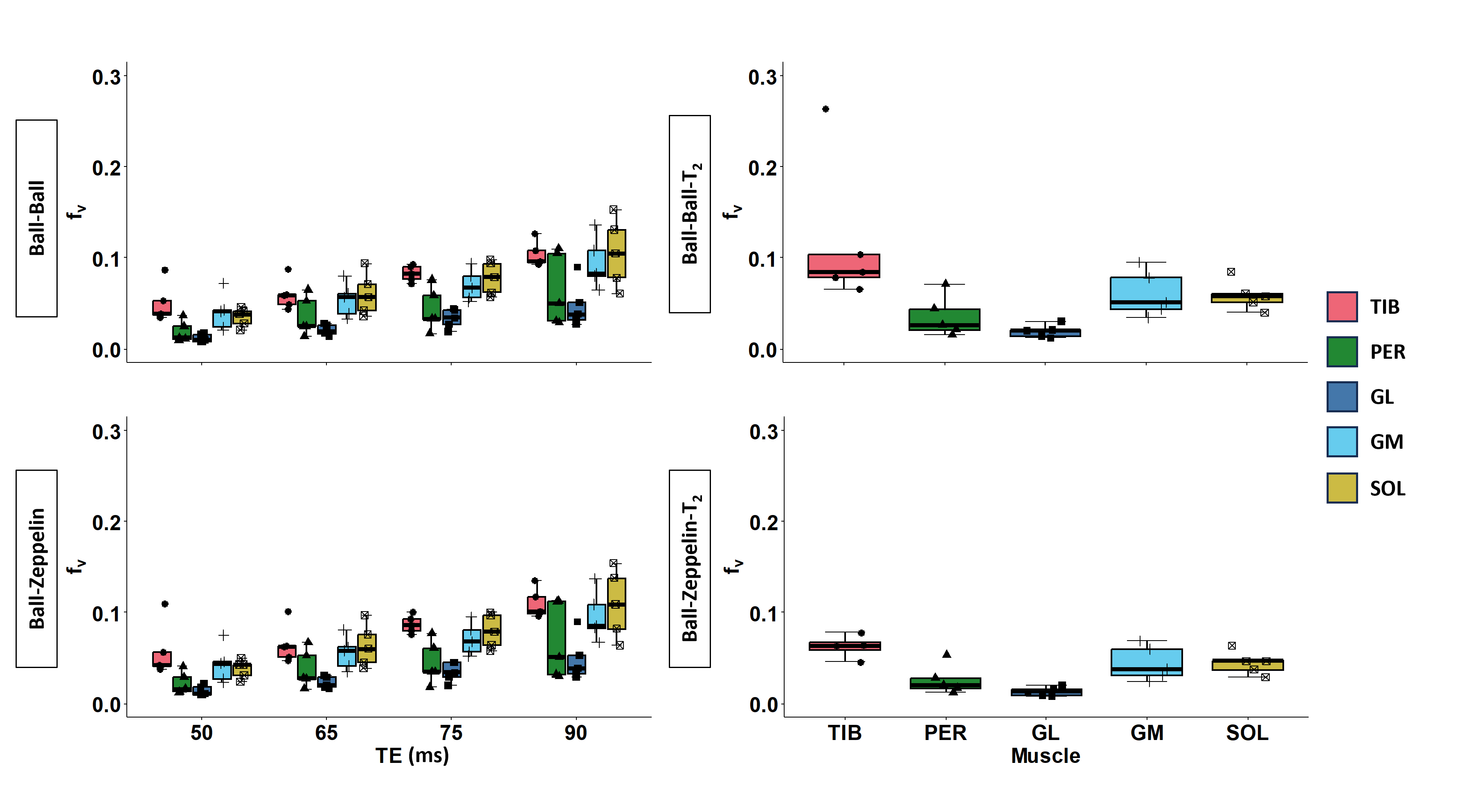}}
\caption{Comparison of vascular fraction across diffusion models with and without $\mathrm{T}_2$ relaxation. Boxplots of ROI-averaged $f_v$ estimates across five calf muscles. The left panels present estimates from models not accounting for $\mathrm{T}_2$ relaxation, illustrating pronounced TE-dependent variations. The right panels show estimates from models incorporating $\mathrm{T}_2$ relaxation.\label{fig5}}
\end{figure*}

As shown in the right-hand column of Figure \ref{fig2}, MD, $D_t$, FA and $f_v$ maps were also effectively generated using diffusion-relaxation models. 

Considering MD and $D_t$, PER and GM showed higher values than TIB across all implemented diffusion–relaxation models. Using TIB as reference, the mean relative differences for these two muscles were approximately +1.1\% to +2.6\% for MD in the Zeppelin-$\mathrm{T}_2$ model, +2.4\% to +7.7\% for $D_t$ in the Ball–Ball-$\mathrm{T}_2$ model, and +10.8\% to +14.1\% for MD in the Ball–Zeppelin-$\mathrm{T}_2$ model. In the Ball–Zeppelin-$\mathrm{T}_2$ model, MD was also higher in GL and SOL than in TIB, with relative differences of approximately +7.3\% and +2.4\%, respectively.

As to FA, Figure \ref{fig4} shows that the SOL muscle exhibited the lowest values among all muscles. Compared with the muscle showing the highest FA values, SOL was lower by up to approximately 42\% in the Zeppelin-$\mathrm{T}_2$ model and 39\% in the Ball–Zeppelin-$\mathrm{T}_2$ model.

Regarding $f_v$, as shown in Figure \ref{fig5}, TIB exhibited the highest values across both two-compartment diffusion–relaxation models. When normalized to TIB, the largest reductions in the other muscles were approximately 84\% in the Ball–Ball-$\mathrm{T}_2$ model and 88\% in the Ball–Zeppelin-$\mathrm{T}_2$ model.

\begin{figure*}
\centerline{\includegraphics[width=0.8\textwidth]{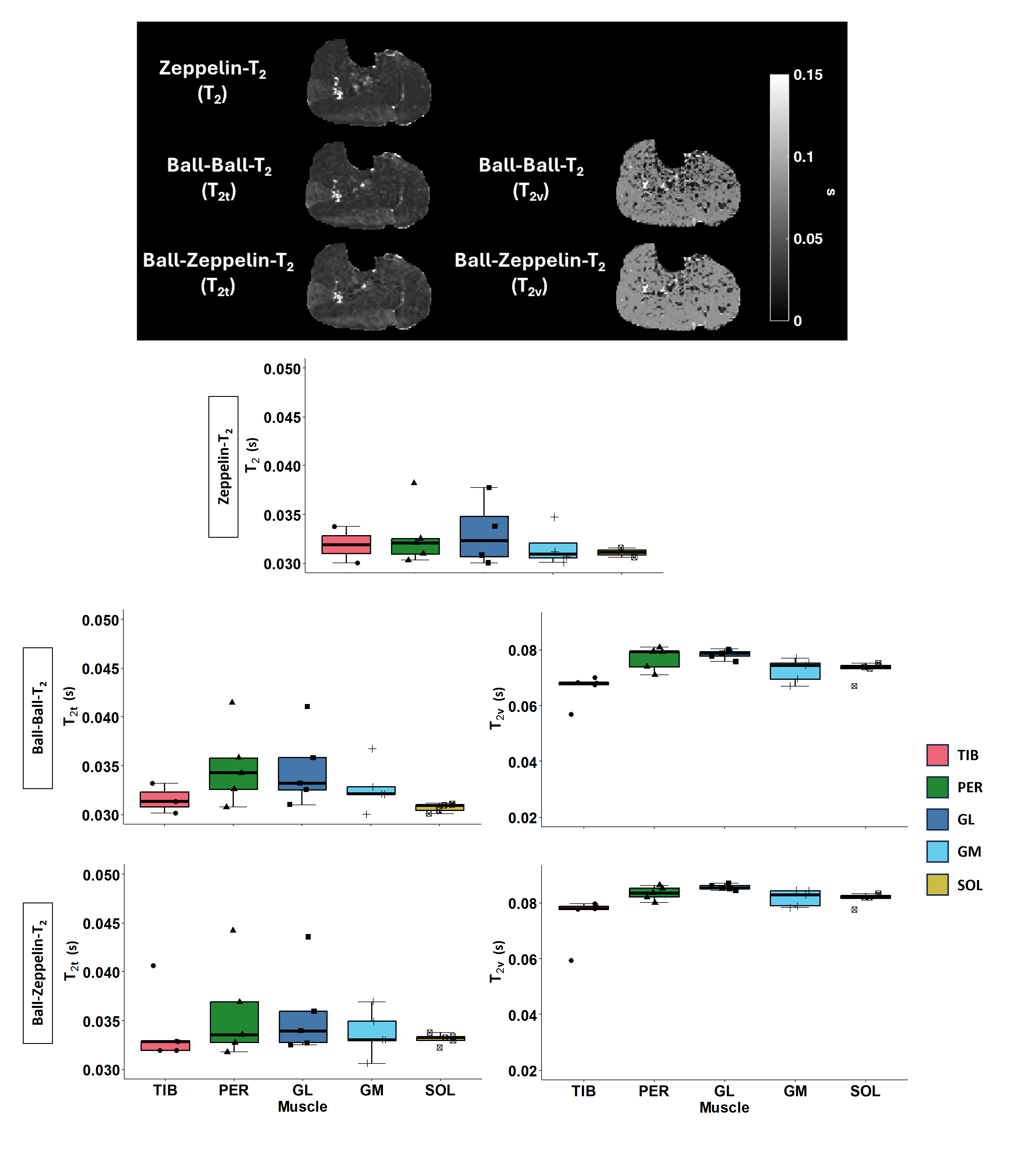}}
\caption{$\mathrm{T}_2$ relaxation times from combined diffusion-relaxation models. Left panel: Representative parametric maps of tissue ($\mathrm{T}_2$t) and vascular ($\mathrm{T}_2$v) relaxation times derived from $\mathrm{T}_2$-extended diffusion models in one subject. Right panels: Boxplots summarizing ROI-averaged $\mathrm{T}_2$ relaxation times across all subjects for five calf muscles.\label{fig6}}
\end{figure*}

Figure \ref{fig6} shows representative maps and boxplots for $\mathrm{T}_2$, $\mathrm{T}_2$t and $\mathrm{T}_2$v estimates from the considered diffusion-relaxation models (Zeppelin-$\mathrm{T}_2$, Ball-Ball-$\mathrm{T}_2$ and Ball-Zeppelin-$\mathrm{T}_2$). Without imposing any explicit constraint, the two-compartment models estimated $\mathrm{T}_2$v as longer than $\mathrm{T}_2$t in almost all voxels. Specifically, mean $\mathrm{T}_2$t values ranged from approximately 31 to 35 ms in the Ball–Ball-$\mathrm{T}_2$ model and from 33 to 36 ms in the Ball–Zeppelin-$\mathrm{T}_2$ model. Mean $\mathrm{T}_2$v values ranged from approximately 66 to 78 ms in the Ball–Ball-$\mathrm{T}_2$ model and from 75 to 86 ms in the Ball–Zeppelin-$\mathrm{T}_2$ model.

Overall, most $\mathrm{T}_2$-related estimates tended to be higher in PER and GL, and often also in GM, than in TIB and SOL. This pattern is compatible with the expected involvement of these muscles in the exercise performed before scanning, although the absence of a pre-exercise baseline prevents a direct interpretation in terms of exercise-induced changes.

\section{Discussion}\label{sec4}

In this paper, we introduced novel two-compartment diffusion-relaxation models for skeletal muscle MRI, showed their preliminary application in healthy subjects, and compared the results with simpler, standard diffusion models that do not account for relaxation effects.

While prior studies have successfully used both $\mathrm{T}_2$ mapping and DTI to monitor SM changes, for example in injury recovery (\cite{monte2023nmrbiomed}), exercise (\cite{keller2020eurad}) and muscle denervation (\cite{ha2015jmri}) these approaches still rely on separate acquisitions and independent modeling of diffusion and relaxation information. In contrast, our work constitutes the first application of an integrated diffusion‑$\mathrm{T}_2$ relaxation model enabling simultaneous acquisition and combined parameter estimation within SM, opening new possibilities for comprehensive and reliable tissue microstructure characterization.

The two-compartment diffusion-relaxation models that we developed estimated longer $\mathrm{T}_2$v than $\mathrm{T}_2$t in almost all the voxels as expected without the need to impose any constraint. The absolute $\mathrm{T}_2$ values predicted in both muscle tissue and vascular compartment are consistent with existing literature (\cite{marty2016nmrbiomed}, \cite{yao2015rheumatology}, \cite{yanagisawa2011jmri}), taking into consideration that the largest majority of studies do not separate tissue and blood contribution, and that blood $\mathrm{T}_2$ has a strong dependence on oxygenation (\cite{lin2012mrmatpbm}, \cite{lu2008mrm}).

Most models showed an increase in $\mathrm{T}_2$ (for the diffusion-relaxation models) and diffusivity in the gastrocnemius lateralis and gastrocnemius medialis - plantarflexors of the ankle joint - and peroneous - a stabilizer -, compared in particular with the tibialis - dorsiflexor and antagonist which is relatively quiet on EMG \cite{sara2021single}. This is consistent with the expected effect of the exercise performed before the scans and could be caused by a combination of increased temperature, blood flow, oxygenation, and eventually inflammation leading to increased intramuscular water (\cite{yanagisawa2011jmri}, \cite{ababneh2008mrmpbm})
It is important to note that all measurements in this study were performed after exercise, without a pre-exercise baseline. This means that our comparisons between muscles reflect post-exercise states rather than relative changes from rest. Consequently, parameters differences cannot be directly interpreted as increases or decreases from a resting condition; they represent the balance between local conditions at rest (in terms of perfusion, intramuscular pressure, temperature etc.) and their changes due to the exercise.
In future work, MRI scans both before and after exercise (or other events of interest) should be performed to disentangle effective changes from intrinsic local variability in the examined biomarkers.

It should also be noted that the temporal dynamics of the mechanisms at play have to be carefully considered before designing the experiments. As dMRI acquisition last at least a few minutes and more complex models require longer acquisition protocols, it is important to make sure that acquisitions are performed in relatively stable conditions and at relevant time points. Simpler models should be favoured to examine very rapid phenomena, and a trade-off between desired complexity and scan duration should always be considered.

\begin{figure*}
\centerline{\includegraphics[height=20cm]{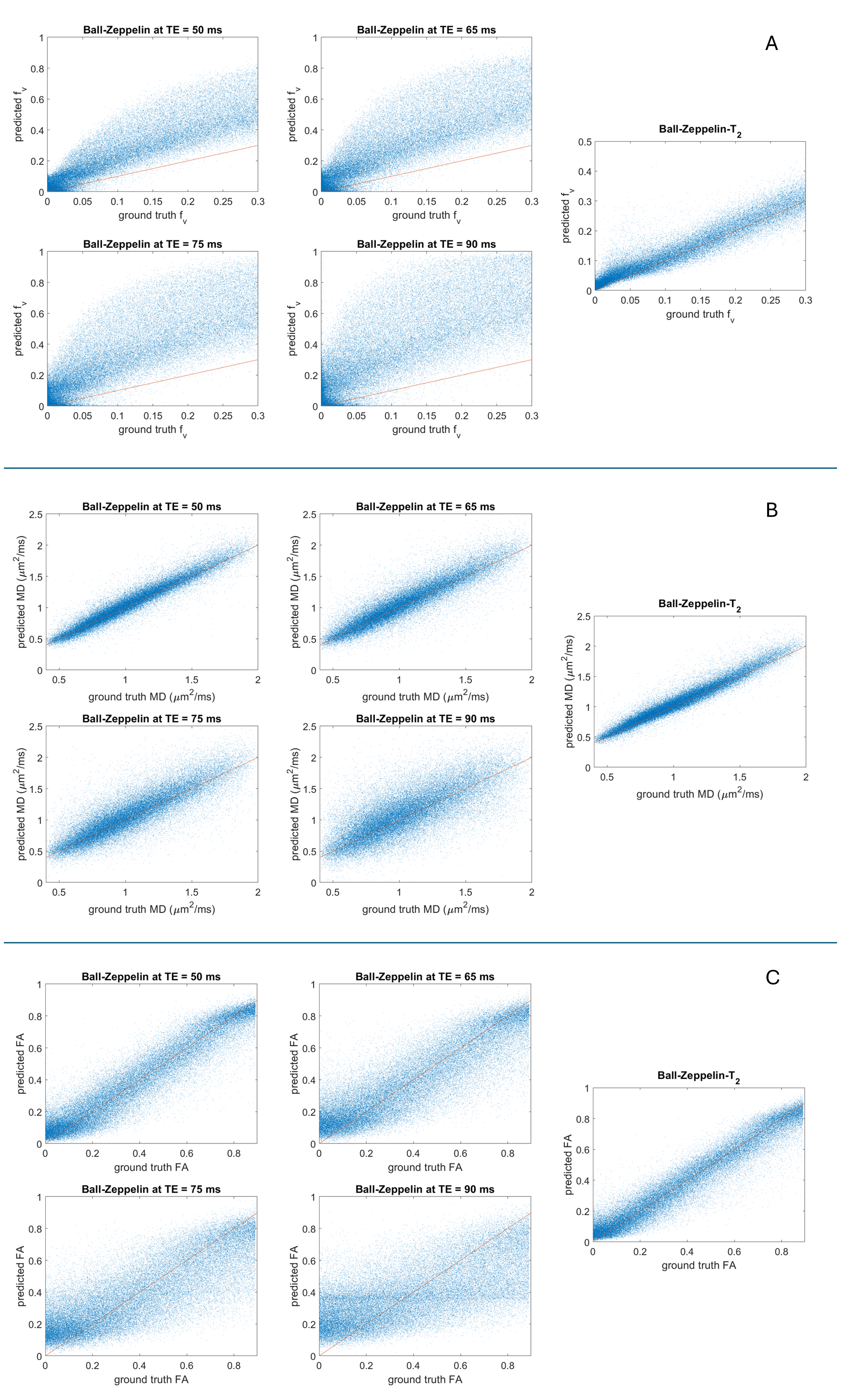}}
\caption{Results from simulations for the estimation of $f_v$ (panel A), MD (panel B) and FA (panel C) with the Ball-Zeppelin and Ball-Zeppelin-$\mathrm{T}_2$ model. Blue dots correspond to estimates from single signals against the corresponding ground truth; the red line is the identity line (perfect estimation).The Ball-Zeppelin model clearly overestimates $f_v$, while the Ball-Zeppelin-$\mathrm{T}_2$ model is much more accurate. MD estimates are much more accurate and just show increased variability for Ball-Zeppelin at longer TEs, but no evident bias. FA estimates show substantially increased variability  for Ball-Zeppelin at longer TEs as well as overestimation of smaller fractions and underestimation of higher fractions. \label{fig7}}
\end{figure*}

We also showed a clear dependence of the estimated parameters on the choice of TE, with all diffusivity metrics decreasing and FA and vascular fractions increasing with longer TE.
To investigate the possible reasons for this dependence, we sampled ground truth parameters uniformly from the ranges in table \ref{table3}, simulated signals in Matlab from $10^6$ combinations of those ground truth parameters using the mathematical formula for the Ball-Zeppelin-$\mathrm{T}_2$ model (equation \ref{eq7}) with the same acquisition protocol as in our real data and adding Rician noise to obtain an SNR of 30 on the b=0 volume with shortest TE. We estimated the parameters of the Ball-Zeppelin-$\mathrm{T}_2$ model from the full signals and the parameters of the Ball-Zeppelin model from the normalised signals at each TE. Results showed a clear overestimation of $f_v$ by the Ball-Zeppelin model (figure \ref{fig7}A), with higher bias and variability at higher TEs. In contrast, the Ball-Zeppelin-$\mathrm{T}_2$ model could estimate $f_v$ with much higher accuracy. Indeed, the correlation coefficient decreases and both bias and error increase from the Ball-Zeppelin-$\mathrm{T}_2$ model to the Ball-Zeppelin model at short TEs to the Ball-Zeppelin model at long TEs (table \ref{table4}).
However, our simulations could not demonstrate the decreasing trend of diffusivity at higher TEs shown in our real data, for example MD was very accurately estimated by the Ball-Zeppelin-$\mathrm{T}_2$ model and also by the Ball-Zeppelin model at shorter TEs, while the Ball-Zeppelin estimates at longer TEs showed more variability, likely due to noise effects, but no evident bias (figure \ref{fig7}B and table \ref{table4}).  The overestimation at longer TEs in the real data is likely due to other effects not considered in our simulations, such as diffusion restriction. It is possible that the diffusion time is automatically set to be higher at longer TEs on the scanner that we used. Unfortunately, we do not have access to the diffusion time set in our experiments, but, if this is the case, we can speculate that the estimated diffusivities decrease with TE due to restriction effects.
As for anisotropy, our simulations showed accurate FA estimation for Ball-Zeppelin-$\mathrm{T}_2$ and for Ball-Zeppelin at the shortest TE, while at longer TEs, Ball-Zeppelin estimates were highly variable and tended to slightly overestimate the ground truth. Indeed, the correlation coefficient decreases and the error increases from the Ball-Zeppelin-$\mathrm{T}_2$ model to the Ball-Zeppelin model at short TEs to the Ball-Zeppelin model at long TEs (table \ref{table4}). The bias also increases, reaching non-negligible values at the longest TEs. However, this trend seems less substantial than in real data, so we can speculate that diffusion time dependence or other unaccounted effects might play a role there.

\begin{table}
    \begin{center}
     \caption{Ranges used for sampling the ground truth values in the simulation experiment}
    \label{table3}
     \begin{tabular}{p{1cm}|p{1.5cm}|p{1.7cm}}
        \textbf{Par} & \textbf{min} & \textbf{max}\\
        \toprule
        $f_v$ & 0 & 0.3\\
        $D_v$ & 10 $\mu m^2/ms$ & 100 $\mu m^2/ms$\\
        $D_{\parallel}$ & 1 $\mu m^2/ms$ & 2 $\mu m^2/ms$\\
        $D_{\perp}$ & 0.1 $D_{\parallel}$ & $D_{\parallel}$\\
        $\theta$ & 0 & 0*\\
        $\phi$ & 0 & 0*\\
        $T_{2v}$ & 60 ms & 150 ms\\
        $T_{2t}$ & 20 ms & 50 ms\\
        \bottomrule
    \end{tabular}
    \end{center}
Note: * The angles were kept at 0 for all simulations, so that the Zeppelin was always oriented in the slice direction, to reduce the number of variables in the experiment without loss of generality.
\end{table}

\begin{table*}
    \centering
     \caption{Agreement, bias and error measures for Ball-Zeppelin and Ball-Zeppelin-$\mathrm{T}_2$ on the simulated signals. r = Pearson's correlation coefficient between predicted and ground truth values; ME = mean error (bias); RMSE = root mean squared error; NRMSE = normalised root mean squared error}
    \begin{tabular}{l l|c|c|c|c}
        & & \textbf{r} & \textbf{ME} & \textbf{RMSE} & \textbf{NRMSE}\\
        \toprule
        & Ball-Zeppelin-$\mathrm{T}_2$ & \textbf{0.952} & \textbf{0.016} & \textbf{0.032} &\textbf{0.308}\\
        \multirow{3}{*}{\textbf{$f_v$}} & Ball-Zeppelin, TE = 50 ms & 0.895 & 0.135 & 0.179 & 1.712\\
        & Ball-Zeppelin, TE = 65 ms & 0.857 & 0.177 & 0.233 & 2.225\\
        & Ball-Zeppelin, TE = 75 ms & 0.829 & 0.209 & 0.272 & 2.594\\
        & Ball-Zeppelin, TE = 90 ms & 0.780 & 0.261 & 0.332 & 3.163\\ [3pt]
        \toprule
        & Ball-Zeppelin-$\mathrm{T}_2$ & \textbf{0.965} & 0.044 ${\mu}m^2/s$ & \textbf{0.101 ${\mu}m^2/s$} & \textbf{0.096}\\
        \multirow{3}{*}{\textbf{MD}} & Ball-Zeppelin, TE = 50 ms & 0.956 & \textbf{0.030 ${\mu}m^2/s$} & 0.108 ${\mu}m^2/s$ & 0.103\\
        & Ball-Zeppelin, TE = 65 ms & 0.907 & 0.038 ${\mu}m^2/s$ & 0.158 ${\mu}m^2/s$ & 0.150\\
        & Ball-Zeppelin, TE = 75 ms & 0.851 & 0.047 ${\mu}m^2/s$ & 0.202 ${\mu}m^2/s$ & 0.192\\
        & Ball-Zeppelin, TE = 90 ms & 0.741 & 0.055 ${\mu}m^2/s$ & 0.267 ${\mu}m^2/s$ & 0.255\\ [3pt] \toprule
        & Ball-Zeppelin-$\mathrm{T}_2$ & \textbf{0.958} & -0.003 & \textbf{0.076} & \textbf{0.196}\\
        \multirow{3}{*}{\textbf{FA}} & Ball-Zeppelin, TE = 50 ms & 0.941 & 0.005 & 0.090 & 0.231\\
        & Ball-Zeppelin, TE = 65 ms & 0.873 & 0.002 & 0.129 & 0.332\\
        & Ball-Zeppelin, TE = 75 ms & 0.798 & \textbf{-0.001} & 0.159 & 0.410\\
        & Ball-Zeppelin, TE = 90 ms & 0.656 & -0.003 & 0.199 & 0.513\\
         \bottomrule
    \end{tabular}
    \label{table4}
\end{table*}

Previous studies have investigated the effect of TE on DTI parameters in the brain, generally finding an increase in the axial diffusivity (or the ADC along peripheral nerves) and a decrease in radial diffusivity (or the ADC perpendicularly to peripheral nerves) at longer TEs and little effect on MD (\cite{qin2009mrm}, \cite{does2000mrm}, \cite{johansson2024invrad}). The contrast with our present findings that all diffusivity parameters decreased at longer TEs confirms our conclusion from the simulation experiments: other factors such as the diffusion time may have a role.
The effect of $\mathrm{T}_2$ itself on DTI parameters has been investigated in the SM, with observed variations explained as mediated mainly by SNR (\cite{froeling2013nmrbiomed}), but a correlation of $\mathrm{T}_2$ and diffusivity changes in exercised muscles has also been observed (\cite{ababneh2008mrmpbm}), hinting at more complex mechanism that involve both diffusion and relaxation properties.
On the other hand, our observation, both in real data and in simulations, that the estimated vascular fraction increases with TE simulations has been confirmed in studies applying the extended IVIM model and comparing it to the standard IVIM both in the healthy liver and in breast cancer (\cite{jerome2016extended}, \cite{egnell2022nmrbiomed}).

Besides the already mentioned lack of a pre-exercise reference and uncertainty about the set diffusion time, the main limitations of this study are the very limited number of subjects, due to the exploratory nature of this study, and the inhomogeneous noise level due to the use of a surface coil. The choice of the coil was the best available in the current clinical setting; however, we observed that the noise was significantly higher in the area further from the surface of the coil, especially in the tibialis muscle. Even though we have denoised the images and accounted for noise in the synthetic signals used to train the MLP models, some of the derived values might be biased and affect the comparisons between muscles.

In future work, we envisage a further study in a relatively small group of subjects using a pre-clinical or a research clinical scanner to control the diffusion time, with a volume coil to get a homogeneous SNR in the images, and multiple time points to assess the real changes in diffusion and relaxation parameters and assess reproducibility. After our proposed method is technically validated with this preliminary experiment, a study in a larger group should be performed, possibly addressing pathological alterations. In particular, applications to neuromuscular diseases (e.g., muscular dystrophies, inflammatory myopathies), metabolic disorders, and rehabilitation interventions would allow assessing the sensitivity of diffusion–relaxation biomarkers to disease progression and therapeutic response.
The diffusion-relaxation models could also be extended by including permeability and/or restriction effects, if this is justified by the experiments.

\section{Conclusions}\label{sec5}

This paper shows how combined diffusion-relaxation MRI represents a powerful non-invasive tool for the comprehensive assessment of muscle microstructure and composition. The study explores its ability to effectively mitigate the confounding effects of $\mathrm{T}_2$ relaxation on diffusion parameters, leading to more accurate quantitative measurements, in particular of the vascular fraction. The preliminary results obtained in healthy subjects following exercise provide evidence of the technique's sensitivity to physiological changes within muscle tissue. Despite the current limitations, particularly the small sample size and the absence of pre-exercise baseline data, the approach holds significant promise for advancing both fundamental muscle research and clinical diagnostics. Its capacity to provide reliable, non-invasive insights into muscle properties positions it as a valuable method for understanding and monitoring conditions ranging from neuromuscular diseases to exercise adaptation and rehabilitation outcomes.

\bmsection*{Author contributions}

Matteo Figini, Paddy Slator, and Alfonso Mastropietro designed the study. Alfonso Mastropietro and Valeria Elisa Contarino performed the data acquisition. Matteo Figini and Alfonso Mastropietro performed data analysis, wrote the draft of the manuscript, and created the figures. Paddy Slator, Valeria Elisa Contarino, Eleftheria Panagiotaki, and Giovanna Rizzo contributed to revising the drafts of the manuscript. All authors approved the final version.

\bmsection*{Acknowledgments}
The authors extend their heartfelt appreciation to Mr. Luciano Lombardi from the Neuroradiology Unit, Fondazione IRCCS Ca'Granda Ospedale Maggiore Policlinico, Milano, Italy, for his technical assistance in conducting MRI scans.

\bmsection*{Financial disclosure}

This work received support from the Royal Society International Exchanges scheme (IEC\textbackslash R2\textbackslash 212086) and from the Italian Ministry of University and Research (MUR) under the program PRIN 2022 (project NINFEA 2022MXW52Y CUP: B53C24006690006).

\bmsection*{Conflict of interest}

The authors declare no potential conflict of interest.

\bibliography{wileyNJD-AMA}

\bmsection*{Supporting information}

Additional supporting information may be found in the
online version of the article at the publisher’s website.

\end{document}